\documentclass[reprint, superscriptaddress, 
floatfix, longbibliography, 
amsfonts, amssymb, amsmath]{revtex4-1}

\usepackage{graphicx}
\usepackage{bm}
\usepackage{dcolumn}
\usepackage{color}

\usepackage[
colorlinks=true, citecolor=blue, urlcolor=blue, linkcolor=blue,
setpagesize=false
]{hyperref}

\begin{document}
\title{
Quantum criticality in the metal-superconductor transition 
of interacting Dirac fermions on a triangular lattice
}

\author{Yuichi Otsuka}
\email{otsukay@riken.jp}
\affiliation{Computational Materials Science Research Team, 
RIKEN Center for Computational Science (R-CCS), 
Kobe, Hyogo 650-0047, Japan}

\author{Kazuhiro Seki}
\affiliation{Computational Materials Science Research Team, 
RIKEN Center for Computational Science (R-CCS), 
Kobe, Hyogo 650-0047, Japan}
\affiliation{SISSA -- International School for Advanced Studies, 
Via Bonomea 265, 34136 Trieste, Italy}
\affiliation{Computational Condensed Matter Physics Laboratory, 
RIKEN, 
Wako, Saitama 351-0198, Japan}

\author{Sandro Sorella}
\affiliation{Computational Materials Science Research Team, 
RIKEN Center for Computational Science (R-CCS), 
Kobe, Hyogo 650-0047, Japan}
\affiliation{SISSA -- International School for Advanced Studies, 
Via Bonomea 265, 34136 Trieste, Italy}
\affiliation{Democritos Simulation Center CNR--IOM Instituto Officina 
dei Materiali, Via Bonomea 265, 34136 Trieste, Italy}

\author{Seiji Yunoki}
\affiliation{Computational Materials Science Research Team,
RIKEN Center for Computational Science (R-CCS), 
Kobe, Hyogo 650-0047, Japan}
\affiliation{Computational Condensed Matter Physics Laboratory,
RIKEN,
Wako, Saitama 351-0198, Japan}
\affiliation{Computational Quantum Matter Research Team, 
RIKEN Center for Emergent Matter Science (CEMS), 
Wako, Saitama 351-0198, Japan}

\date{\today}

\begin{abstract}
 We investigate a semimetal-superconductor phase transition of
 two-dimensional Dirac electrons at zero temperature by large-scale and 
 essentially unbiased quantum Monte Carlo simulations for the
 half-filled attractive Hubbard model on the triangular lattice, in the
 presence of alternating magnetic $\pi$ flux, that is introduced to
 construct two Dirac points in the one-particle bands at the Fermi
 level.
 This phase transition is expected to describe quantum criticality of
 the chiral XY class in the framework of the Gross-Neveu model, where,
 in the ordered phase, the $U(1)$ symmetry is spontaneously broken and 
 a mass gap opens in the excitation spectrum.
 We compute the order parameter of the $s$-wave superconductivity and
 estimate the quasiparticle weight from the long-distance behavior of the
 single-particle Green's function. 
 These calculations allow us to obtain the critical exponents of this
 transition in a reliable and accurate way.
 Our estimate for the critical exponents is in good agreement with those
 obtained for a transition to a Kekul\'{e} valence bond solid, 
 where an emergent $U(1)$ symmetry is proposed 
 [Z.-X. Li \textit{et al.}, Nat. Commun. {\bf 8}, 314 (2017)].
\end{abstract}

\maketitle

\section{\label{sec:intro}Introduction}

The Hubbard model with the Dirac dispersion has been intensively
investigated in recent years, since it represents an ideal platform to
study interaction-driven quantum phase transitions in a controllable way.
Several decades ago, it was realized that the Hubbard model on the
honeycomb lattice, the canonical lattice model for the interacting Dirac
electrons, can be investigated with an unbiased numerical method,
showing a quantum phase transition between a semimetal and 
an antiferromagnetic insulating phase at a finite value of the critical 
interaction strength, $U_{\textrm{c}}/t>0$~\cite{Sorella_EPL1992}. 
This is in contrast to the case of the square lattice, where the nesting 
instability yields a trivial $U_{\textrm{c}}/t=0$~\cite{Hirsch_PRB1985}.
This subject has attracted much attention especially after two different 
spin liquid 
phases were proposed in the Hubbard model on the honeycomb and 
the $\pi$-flux square lattices~\cite{Meng_Nature2010,Chang_PRL2012}.
Although subsequent studies have concluded that the spin liquid phases 
in these models are 
unlikely~\cite{Sorella_SR2012,Hassan_PRL2013,Assaad_PRX2013,Ixert_PRB2014,ParisenToldin_PRB2015,Otsuka_PRX2016},
this query has also been an opportunity to investigate the
interaction-driven phase transition of the Dirac fermions by modern
numerical and analytical techniques with a renewed interest, focusing on
fermionic quantum criticality and universality classes.

Herbut and co-workers argued that the quantum criticality of the
semimetal-antiferromagnetic transition of the Dirac fermions is
described by the Gross-Neveu (GN)
model~\cite{Gross_PRD1974,Herbut_PRL2006,Herbut_PRB2009a}.
More generally, in the framework of the GN theory, 
it is known that there are three universality classes depending on 
the symmetries of the order parameters in the ordered phases, 
where, in all cases, the chiral symmetry is broken and a finite gap 
appears in the excitation spectrum~\cite{Rosenstein_PhysLettB1993}.
Since this behavior is universal, it should be detected also in
Dirac fermions lattice models  relevant for condensed matter
physics.
Indeed, the universal quantum criticality of the
semimetal-antiferromagnetic transition, i.e., the $SU(2)$ symmetry
breaking, corresponding to the chiral Heisenberg class in terms of the
GN model, was numerically confirmed by calculating the critical
exponents for the Hubbard model on the honeycomb lattice and on the
square lattice with 
$\pi$ flux~\cite{Assaad_PRX2013,Ixert_PRB2014,ParisenToldin_PRB2015,Otsuka_PRX2016}.
On the other hand, the chiral Ising class, which describes the $Z_{2}$
symmetry breaking in the interacting Dirac fermions, was examined in the
charge-density-wave (CDW) transition of the spinless $t$-$V$ model on
the same lattices~\cite{Wang_NJP2014,Li_NJP2015,Hesselmann_PRB2016,Wang_PRB2016}.
In both cases, quantum Monte Carlo (QMC) methods, 
designed for the simple Hubbard-like lattice models,
enable us to obtain the critical exponents with high accuracy,
which was difficult by analytical methods based on
renormalization-group (RG) approaches~\cite{Rosenstein_PhysLettB1993,Gracey_NuclPhysB1990,Gracey_1994,Vasilev_1993,Herbut_PRB2009a,Rosa_PRL2001,Hofling_PRB2002,Janssen_PRB2014}.

Among the three universality classes categorized by the GN theory, the
remaining one, i.e., the chiral XY class corresponds to the $U(1)$
symmetry breaking.
For this class, the QMC results have been obtained on the basis of 
the Kekul\'{e} valence-bond-solid (VBS) transition~\cite{Li_NatCom2017,Xu_arXiv2018}.
Although the Kekul\'{e} VBS state is naively understood as a consequence
of the $Z_{3}$ symmetry breaking, the RG arguments predict that the
$U(1)$ symmetry emerges at the quantum critical point because of peculiar
gapless fermion fluctuations, implying that the Kekul\'{e} VBS
transition belongs to the chiral XY 
class~\cite{Li_NatCom2017,Scherer_PRB2016,Zerf_PRD2017,Classen_PRB2017}.
The emergent $U(1)$ symmetry was indeed observed in the QMC
simulations~\cite{Li_NatCom2017,Xu_arXiv2018}.
However, since it occurs only at the critical point,
the scaling region can be significantly narrow, which may affect
the QMC estimates of the critical exponents.
Therefore, 
it is desirable to provide an independent estimate
based on a lattice model where only the $U(1)$ symmetry is present.
Moreover the fermion anomalous dimension has not been obtained by the QMC method, yet.

In this paper, we study the quantum criticality of the chiral XY class
on the basis of a lattice model which directly exhibits the $U(1)$
symmetry breaking.
Specifically, the attractive Hubbard model on the triangular lattice
with alternating $\pi$ flux at half electron filling is investigated by large-scale QMC simulations.
In this model, the attractive on-site interaction drives the semimetal
phase, which is stable in the weak-coupling region, to the $s$-wave
superconducting (SC) phase where the $U(1)$ symmetry is broken. 
We calculate the order parameter of the $s$-wave SC phase with high
accuracy on lattices containing up to 2500 sites.
The phase transition is also examined from the long-distance
behavior of the single-particle Green's function, from which the quasiparticle weight
is estimated.
The critical exponents are obtained by a careful finite-size scaling 
analysis applied to these quantities. 
The critical point $U_{\mathrm{c}}/t$ and the correlation-length
exponent $\nu$, estimated independently from the SC order
parameter and the quasiparticle weight, show a good agreement, 
which clearly supports the quality of our calculation.

The rest of the paper is organized as follows.
In the next section, we define the model and briefly explain the
simulation method.
The results of the SC order parameter and the quasiparticle weight are 
shown in Sec.~\ref{sec:result}.
The obtained exponents are compared with the previous results before 
concluding the paper in Sec.~\ref{sec:conclusion}.

\section{\label{sec:model}Model and Method}

\subsection{Model}

\begin{figure}[ht]
 \centering
 \includegraphics[width=0.23\textwidth]{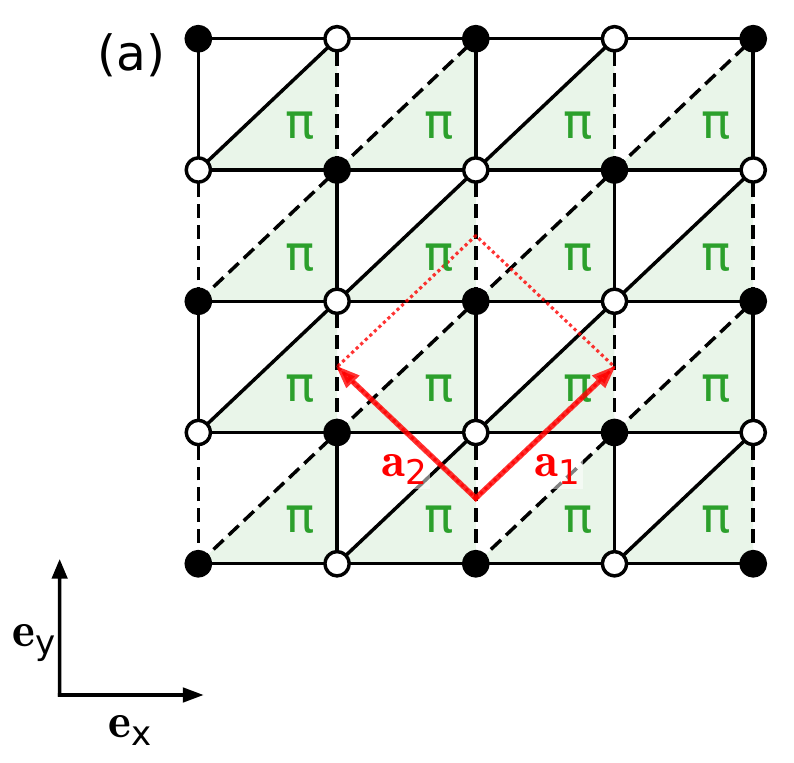}
 \includegraphics[width=0.23\textwidth]{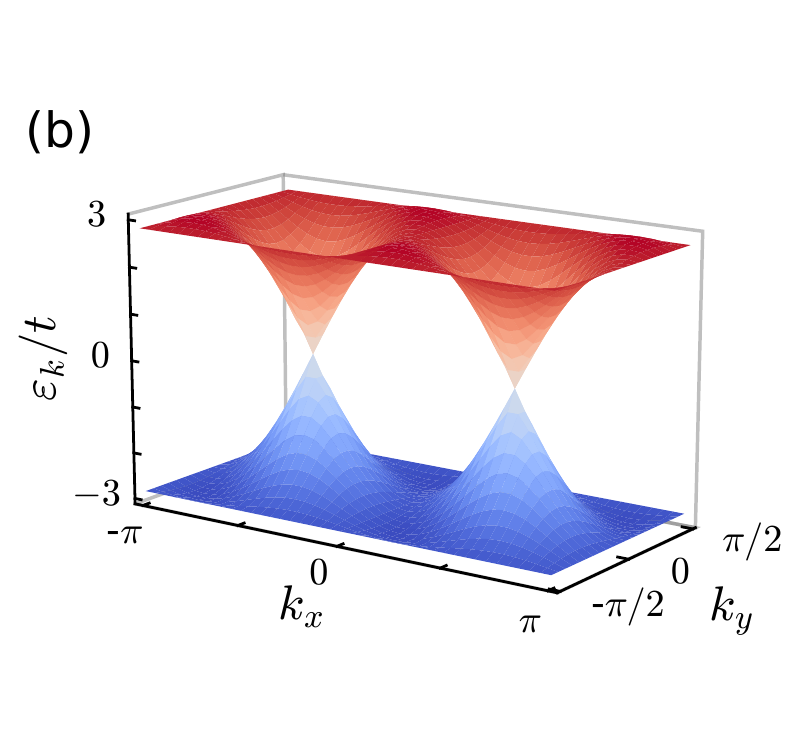}
 \caption{\label{fig:lattice}%
 (a) Lattice structure of the triangular lattice with alternating
 $\pi$ flux.
 Each unit cell spanned by the primitive translational vectors
 (red arrows), ${\bm a}_{1}=(1,1)$ and ${\bm a}_{2}=(-1,1)$, contains two sites 
 indicated by solid and open circles, where the lattice constant is set to be one. 
 ${\bm e}_x=(1,0)$ [${\bm e}_y=(0,1)$] is the unit vector along the $x$ ($y$) direction. 
 Transfer integrals for solid and dashed lines
 are $-t$ and $-te^{i\pi}$, respectively.
 The phase factor for the dashed line leads to
 a magnetic $\pi$ flux penetrating through each shaded triangle.
 (b) Energy dispersion in the noninteracting limit ($U/t=0$).
 The Fermi level is located at $\varepsilon_{\bm{k}}/t=0$ for half filling.
 }
\end{figure}

The attractive Hubbard model is expressed by the following Hamiltonian:
\begin{equation}
 H  = 
\sum_{\langle i,j\rangle,\sigma} t_{ij}
 \left( c_{i \sigma}^{\dagger}c_{j \sigma}
 + \text{h.c.} \right) - U \sum_{i} n_{i \uparrow  }  n_{i \downarrow},
 \label{eq:Hamiltonian}
\end{equation}
where $c_{i \sigma}^{\dagger}$ creates an electron with 
spin $\sigma(=\uparrow,\downarrow)$ at site $i$, 
located at ${\bm r}_i=i_x{\bm e}_x+i_y{\bm e}_y$ 
on the triangular lattice ($i_x$ and $i_y$: integer) 
[see Fig.~\ref{fig:lattice}(a)], and 
$n_{i \sigma}=c_{i \sigma}^{\dagger}c_{i \sigma}$. 
The sum indicated by $\langle i,j\rangle$ runs over all pairs of
neighboring sites $i$ and $j$ on the triangular lattice.  
The first term represents the kinetic energy 
defined by the tight-binding model with
the transfer integrals $t_{ij}=-|t_{ij}|e^{i \theta_{ij}}$. 
We consider the model with 
uniform $|t_{ij}|=t$ 
for both links of the triangular lattice, as depicted by solid and
dashed lines in Fig.~\ref{fig:lattice}(a). 
The magnetic $\pi$ flux is imposed
for every other triangle, which is realized by choosing
$\theta_{ij}=\pi$ for the dashed links.
Owing to this flux pattern,
the noninteracting energy dispersion,
$\varepsilon_{\bm{k}}^{\pm}=\pm 2t \sqrt{1 + \cos^{2}\left(k_{x} + k_{y} \right) + \cos^{2}k_{x} - \cos^{2}k_{y}}$,
has two Dirac points at $\bm{K}=(\pm\pi/2,0)$,
as shown in Fig.~\ref{fig:lattice}(b).
The second term in Eq.~(\ref{eq:Hamiltonian}) represents
the attractive interaction ($U>0$), which induces the
phase transition to the $s$-wave SC phase with increasing $U$.
We study the model at half filling where the Fermi level
is located at the Dirac points ($\varepsilon_{\bm{k}=\bm{K}}^{\pm}=0$); 
thus the low-laying excitations are described by the spin-$1/2$ Dirac
fermions~\footnote{The spin liquid phase was found in the repulsive model on the 
same lattice by variational cluster approximation~\cite{Rachel_PRL2015}.}.

In contrast to the $\pi$-flux model on the square
lattice~\cite{Affleck_PRB1988a,Affleck_PRB1988b,Otsuka_PRB2002,Ixert_PRB2014,ParisenToldin_PRB2015,Otsuka_PRX2016},
the sublattice symmetry is absent in our model because of the
triangular-lattice geometry.
This indicates that the model (\ref{eq:Hamiltonian}) itself does not
satisfy the chiral symmetry.
In the noninteracting limit, 
the effective low-energy Hamiltonian is obtained by a linear  expansion 
in $\delta \bm{k}=\bm{k}-\bm{K}=(\delta k_{x}, \delta k_{y})$, as  
\begin{align}
 H_{\mathrm{eff}} =
  \pm 2 t \delta k_{x} \sigma_{x}
 +    2 t \delta k_{y} \sigma_{y}
  \mp 2 t \left( \delta k_{x} + \delta k_{y} \right) \sigma_{z},
 \label{eq:Heff}
\end{align}
where $\bm{\sigma}=(\sigma_{x}, \sigma_{y},\sigma_{z})$ represent the
Pauli matrices.
We notice that the usual chiral operator $\sigma_{z}$ does not
anticommute with the effective Hamiltonian,
$\{H_{\mathrm{eff}}, \sigma_{z}\} = H_{\mathrm{eff}} \sigma_{z} + \sigma_{z}H_{\mathrm{eff}}  \neq 0$, 
since Eq.~(\ref{eq:Heff}) has the nonzero $\sigma_{z}$ term.
However, we can still define a general chiral operator as 
$\gamma=\bm{n}_{\gamma} \cdot \bm{\sigma}$ with $\gamma^{2}=\sigma_{0}$,
where $\sigma_{0}$ is a $2 \times 2$ unit operator,
satisfying the condition of the chiral symmetry
at the Dirac points~\cite{Hatsugai_JPCS2011}.
For the specific case of Eq.~(\ref{eq:Heff}),
it is easily shown that 
$\{H_{\mathrm{eff}}, \gamma\} = 0$ holds
for a choice of $\bm{n}_{\gamma}=(\pm1, 1, \pm1)/\sqrt{3}$,
which implies that 
the chiral symmetry is retained in the vicinity of the Dirac points in
the continuum limit.
We thus expect that the critical behavior near the phase transition
is effectively described by the GN theory, because 
only the Dirac dispersion near the Fermi level is relevant
in the low-energy limit of the model~(\ref{eq:Hamiltonian}).

On the other hand, depending on the presence or absence of the
sublattice symmetry, different symmetries are broken at strong
coupling. 
On the square lattice at half filling,
the SC phase transition is accompanied by the CDW transition 
resulting in the $SU(2)$ symmetry breaking~\cite{Scalettar_PRL1989,Moreo_PRL1991},
whereas only the SC phase transition,
i.e., the $U(1)$ symmetry breaking, occurs on the triangular
lattice~\cite{DosSantos_PRB1992,DosSantos_PRB1993}.

We study the model (\ref{eq:Hamiltonian}) from a theoretical point
of view without considering its origin or a possible candidate for
material realization.
However, let us point out that, owing to recent technological
developments in manipulating atoms trapped in optical lattices,
it has become possible to introduce the staggered $\pi$ flux in
triangular lattices~\cite{Struck_PRL2012}.
Since interactions between atoms in the optical lattices are also
tunable~\cite{Hackermuller_Science2010}, 
the transitions studied in this work might be experimentally explored in
the near future.

\subsection{Method}

Since the attractive Hubbard model is free from the negative sign
problem~\cite{Hirsch_PRB1985}, 
we adopt the ground-state projection within the auxiliary-field QMC
method~\cite{Blankenbecler_PRD1981,White_PRB1989}.
In this technique, 
an expectation value of a physical observable $O$ is calculated as
\begin{equation}
 \langle O \rangle =
  \lim_{\tau \rightarrow \infty}
  \frac
  {
  \langle \psi_{\mathrm{L}} |  
  e^{-\frac{\tau}{2} H}
  O
  e^{-\frac{\tau}{2} H}
  | \psi_{\mathrm{R}} \rangle
  }
  {
  \langle \psi_{\mathrm{L}} |  
  e^{-\tau H}
  | \psi_{\mathrm{R}} \rangle
  },
\end{equation}
where $\tau$ is a projection time and 
$\langle \psi_{\mathrm{L}} |$ 
($| \psi_{\mathrm{R}} \rangle$) is a left (right) trial wave function
having a finite overlap with the ground state.
The Suzuki-Trotter decomposition~\cite{Suzuki_1976,Trotter_1959} is
applied to the projection operator,
$e^{- \Delta \tau H} = 
 e^{- \frac{1}{2} \Delta\tau H_{0}}
 e^{-             \Delta\tau H_{1}}
 e^{- \frac{1}{2} \Delta\tau H_{0}}
+\mathcal{O}(\Delta\tau^{3})$,
where $H_{0}$ ($H_{1}$) represents the first (second) term
in Eq.~(\ref{eq:Hamiltonian}), 
and $\Delta \tau = \tau /  N_{\tau}$ with $N_{\tau}$ being integer.
In practice, $\tau$ and $N_{\tau}$ are not infinite but  
set large enough to have negligible systematic errors.
In addition, since the QMC simulations are performed on finite-size
clusters, the results are always affected by finite-size effects.
This error can be systematically eliminated by performing 
simulations on large clusters
and employing well established finite-size scaling analysis, 
as will be shown in the following section.  
The finite-size clusters used in this study are determined by two
orthogonal lattice vectors $(L_{x}\bm{e}_{x}, L_{y}\bm{e}_{y})$ with the same
length $L_{x}=L_{y}=L$ containing $N=L^{2}$ sites [see Fig.~\ref{fig:lattice}(a)],
and the periodic-boundary conditions are imposed along the $x$ and $y$ directions.
On each cluster, we confirm that
the systematic errors are sufficiently small compared to
the statistical errors by choosing 
$\tau = (L+4)/t$ and $\Delta\tau=0.1/t$.

\section{\label{sec:result}Results}

In this section, we first confirm that the CDW order does not develop,
while the SC order does in the strong coupling region.
The $s$-wave SC order parameter is obtained from a simple extrapolation
of the correlation function as a function of $1/L$, which gives a rough 
estimation of the phase boundary $U_{\mathrm{c}}/t$ and the critical 
exponent for the order parameter $\beta$.
Then, by using a standard finite-size scaling analysis, we estimate
$U_{\mathrm{c}}/t$ and $\beta$ more accurately and also obtain the
correlation-length exponent $\nu$.
We then examine the phase transition from the semimetallic region.
The quasiparticle weight is evaluated from the long-distance behavior of 
the single-particle Green's function, which yields another independent
estimation of $U_{\mathrm{c}}/t$ and $\nu$, and also the remaining independent 
exponent $\eta_{\psi}$.

\subsection{\label{appsec:mcdw} Absence of the CDW order}

Before discussing the numerical results,
let us consider our model in the strong-coupling limit by a simple argument.
The attractive Hubbard model is mapped onto the repulsive model under 
a partial (spin-down) particle-hole
transformation~\cite{Shiba_PTP1972,Emery_PRB1976},
\begin{align}
 c_{i \uparrow} & \rightarrow \tilde{c}_{i \uparrow}\\
 c_{i \downarrow}  & \rightarrow \tilde{c}_{i \downarrow}^{\dagger}
 = (-1)^{s_i} c_{i \downarrow},
\end{align}
where $s_i=i_x+i_y$ and hence $(-1)^{s_i}$ yields $1$ ($-1$) for sites
depicted by open (solid) circles in Fig.~\ref{fig:lattice}(a).
Since the triangular lattice is not bipartite, the kinetic term is not
invariant under this transformation:
In the mapped Hamiltonian,
the signs of transfer integrals along the diagonal bonds 
in the square lattice representation, as shown in Fig.~\ref{fig:lattice}(a), 
are all spin dependent~\cite{DosSantos_PRB1993}.
Therefore, in the strong-coupling limit, the effective spin model for the 
mapped repulsive model has an XXZ-type anisotropy, 
which is given by
\begin{equation}
 \tilde{H}_{\mathrm{eff}} = \sum_{\langle i,j\rangle}
 \left\{
   J_{ij}^{z}  S_{i}^{z} S_{j}^{z} +
   J_{ij}^{xy} \left( S_{i}^{x} S_{j}^{x} + S_{i}^{y} S_{j}^{y} \right)
 \right\},
  \label{eq:XXZ}
\end{equation}
where $S_{i}^{\alpha}$ is the spin-$1/2$ operator of the localized
electron described by $\tilde{c}_{i \sigma}$ and $J_{ij}^{\alpha}$
denotes the neighboring spin interaction due to the second-order kinetic
processes.
Since $J_{ij}^{z}=4t^{2}/U \equiv J >0$ for all the bonds in each triangle, 
the antiferromagnetic order in the $z$ direction, which corresponds to
the CDW order in the language of the original attractive Hubbard model, 
is strongly suppressed because of the geometrical frustration.

On the other hand, the order in the $xy$ plane, corresponding to
$s$-wave SC order in the attractive Hubbard model, is not prevented, 
because $J_{ij}^{xy}=J$ between the nearest-neighboring sites and 
$J_{ij}^{xy}=-J$ for the second-neighboring diagonal bond in the square
lattice representation [Fig.~\ref{fig:lattice}(a)].
Therefore, we expect only the $s$-wave SC order, i.e., the $U(1)$
symmetry breaking, in the strong-coupling region, 
and thus the model studied here is particularly appropriate to examine 
the genuine $U(1)$ universality class of the GN transition.

\begin{figure}[ht]
 \centering
 \includegraphics[width=0.45\textwidth]{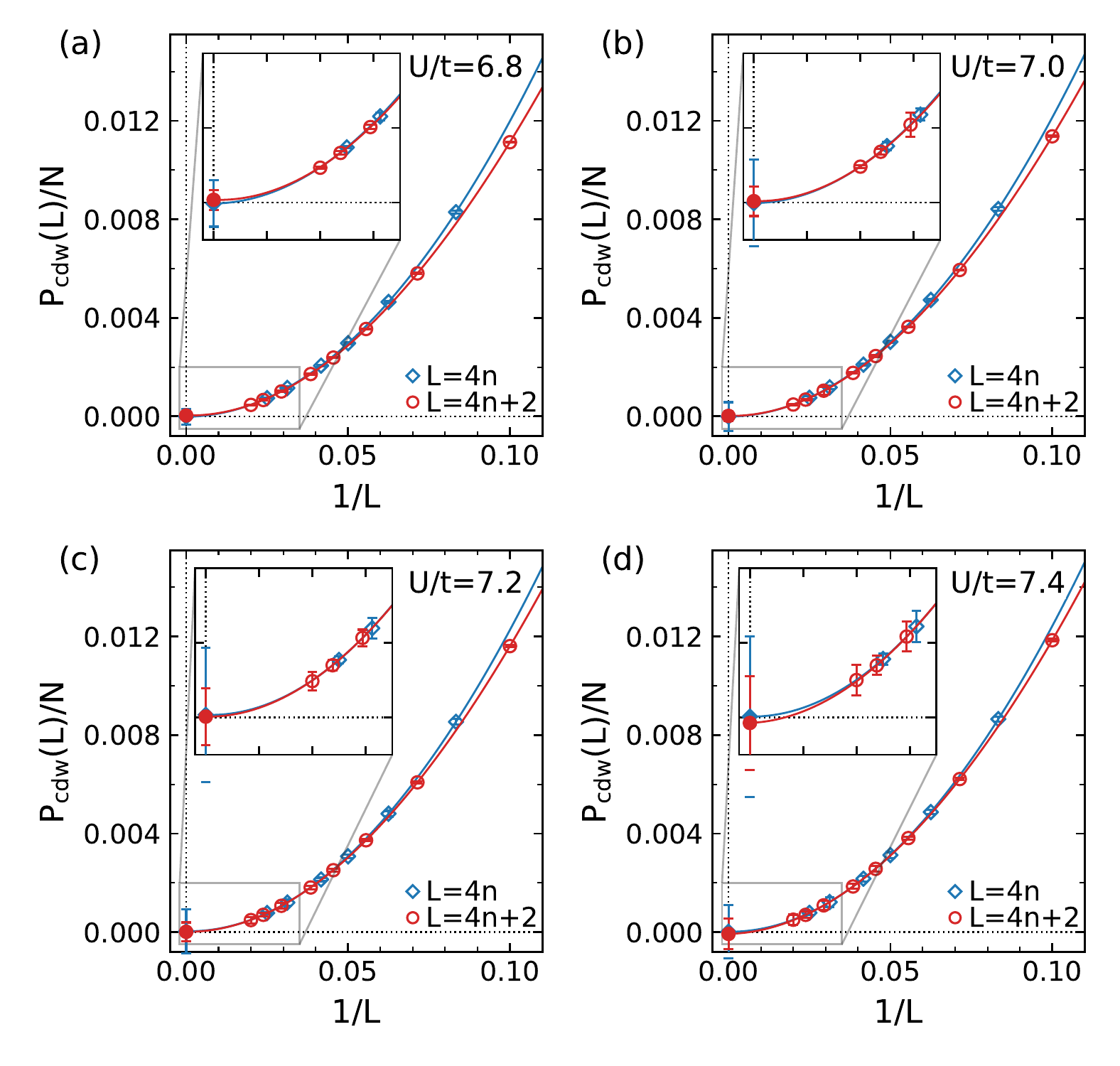}
 \caption{\label{fig:mcdw_poorman_L}%
 CDW correlation function $P_{\mathrm{CDW}}(L)$ 
 extrapolated to the thermodynamic limit for 
 (a) $U/t=6.8$, (b) $U/t=7.0$, (c) $U/t=7.2$, and (d) $U/t=7.4$.
 Open diamonds (circles) represent results for $L=4n$ ($L=4n+2$).
 Solid curves are least-square fits to the data with cubic polynomials
 in $1/L$. Insets show enlarged plots for large $L$.
 The extrapolated values in the thermodynamics limit are shown by filled
 symbols at $1/L=0$.
 }
\end{figure}

We numerically confirm the above argument by directly calculating the CDW
correlation function,
\begin{equation}
 P_{\mathrm{CDW}}(L) = \frac{1}{N} \sum_{i,j}
(-1)^{s_i+s_j}
\langle
\left( n_{i \uparrow} + n_{i \downarrow}\right)
\left( n_{j \uparrow} + n_{j \downarrow}\right)
\rangle,
\label{eq:CDW}
\end{equation}
from which the CDW order parameter is obtained as 
$m_{\mathrm{CDW}}=\lim_{L \rightarrow \infty} \sqrt{P_{\mathrm{CDW}}(L)/N}$.
We define two sequences of clusters, $L=4n$ and $L=4n+2$,
with $n$ being integer.
The difference is that, with the periodic-boundary conditions,
the former has the Dirac points among the allowed momenta and
the latter does not. 
The system sizes studied are 
$L =  8, 12, 16, 20, 24, 32, 40, 48$ 
(this sequence of clusters denoted in the following 
as ``L$_{\mathrm{4n}}$'') and
$L = 10, 14, 18, 22, 26, 34, 42, 50$ 
(this sequence of clusters denoted in the following 
as ``L$_{\mathrm{4n+2}}$''), respectively.
As shown in Fig.~\ref{fig:mcdw_poorman_L}, the CDW order parameter
clearly vanishes in both sequences of clusters L$_{\mathrm{4n}}$ and
L$_{\mathrm{4n+2}}$.

\subsection{\label{subsec:SC}SC order parameter}

\begin{figure}[ht]
 \centering
 \includegraphics[width=0.45\textwidth]{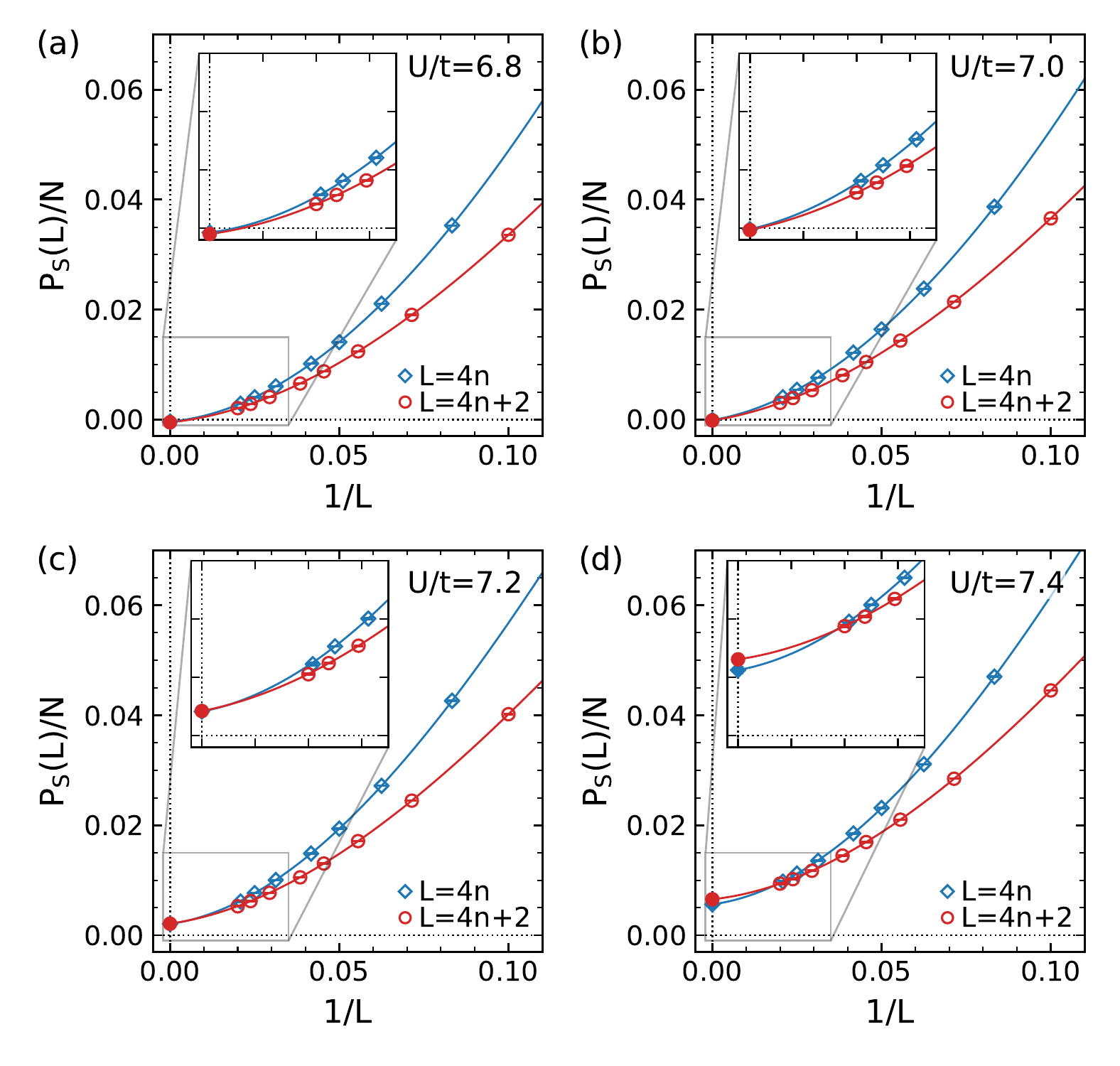}
 \caption{\label{fig:Ps_poorman_L}%
 $s$-wave pairing correlation function
 $P_{\mathrm{s}}(L)$ extrapolated to the thermodynamic limit for 
 (a) $U/t=6.8$, (b) $U/t=7.0$, (c) $U/t=7.2$, and (d) $U/t=7.4$.
 Open diamonds (circles) represent results for $L=4n$ ($L=4n+2$).
 Solid curves are least-square fits to the data with cubic polynomials
 in $1/L$. Insets show enlarged plots for large $L$.
 The extrapolated values are shown by filled symbols at $1/L=0$.
 }
\end{figure}

Next, we calculate the $s$-wave pairing correlation function,
\begin{equation}
 P_{\mathrm{s}}(L) = \frac{1}{N} \sum_{i, j}
  \langle
  c_{i \uparrow}^{\dagger} c_{i \downarrow}^{\dagger}
  c_{j \downarrow}           c_{j \uparrow}
  +
  c_{j \downarrow}           c_{j \uparrow}
  c_{i \uparrow}^{\dagger}   c_{i \downarrow}^{\dagger}
  \rangle,
\end{equation}
on each finite-size cluster of the two sequences L$_{\mathrm{4n}}$ and
L$_{\mathrm{4n+2}}$.
As shown in Fig.~\ref{fig:Ps_poorman_L}, these two sequences indeed show
different finite-size behaviors especially for small $L$ but converge
to consistent values in the thermodynamics limit.
Because of this different asymptotic behavior, 
it is not possible to apply the finite-size scaling analysis
for the mixed data set with L$_{\mathrm{4n}}$ and L$_{\mathrm{4n+2}}$.
One may consider that the suitable choice should be L$_{\mathrm{4n}}$;
it has the Dirac points in the allowed momenta, which is expected to
be important in the long wavelength limit.
However, in the practical QMC simulations,
the sequence L$_{\mathrm{4n+2}}$ has the considerable
advantage that the noninteracting model has a nondegenerate ground
state $|\psi_{U=0}\rangle$, 
which is known as ``closed-shell condition''~\cite{Sorella_PRB2015}.
This is particularly useful because one can select the trial wave
functions as 
$|\psi_{\mathrm{L}}\rangle=|\psi_{\mathrm{R}}\rangle=|\psi_{U=0}\rangle$.
This accelerates the convergence to the limit of $\tau \to \infty$  
(see discussion in Ref.~\onlinecite{Sorella_SR2012})
with much less computational effort 
since the spin-up determinant and the spin-down determinant are
exactly the same in this case, and one needs to compute only one of them
and with half the dimension corresponding to the L$_{\mathrm{4n}}$ case, 
allowing at least four times more efficient computations. 
For this reason, in the rest of the paper, we apply the finite-size
scaling analysis only to the sequence of clusters with $L=4n+2$, 
for which we can obtain 
statistically accurate results up to the largest cluster of $L=50$.

\begin{figure}[ht]
 \centering
 \includegraphics[width=0.30\textwidth]{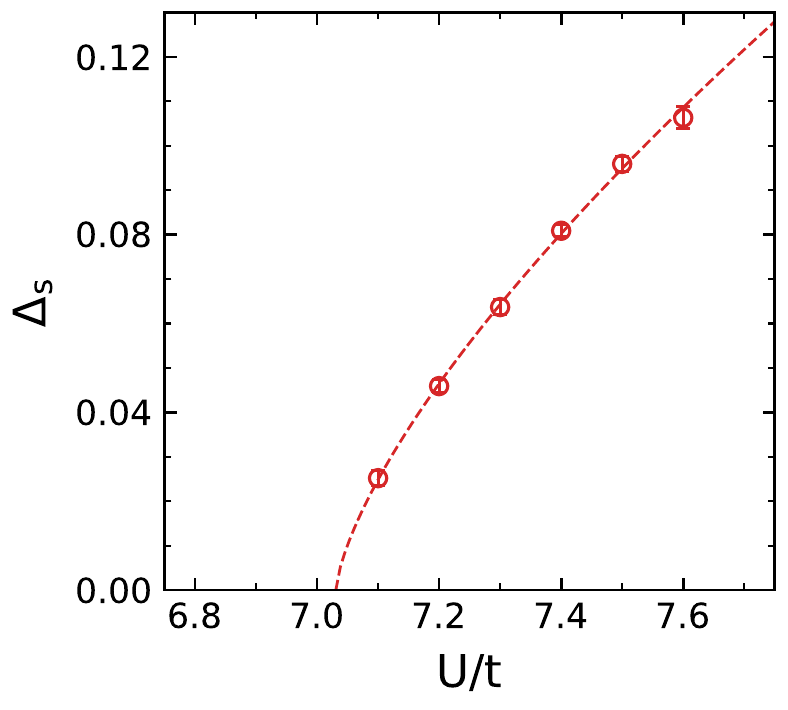}
 \caption{\label{fig:Ps_poorman_U}%
 $s$-wave SC order parameter $\Delta_{\mathrm{s}}$ as a function of $U/t$ 
 for $L=4n+2$.
 A dashed line is a least-square fit to the data points
 with the form of $\Delta_{\mathrm{s}} \sim (U - U_{\mathrm{c}})^{\beta}$,
 from which the critical point and the exponent are estimated as
 $U_{\mathrm{c}}/t=7.03(2)$ and $\beta=0.70(6)$.
 }
\end{figure}

The order parameter $\Delta_{\mathrm{s}}$ of the $s$-wave SC phase is obtained 
by extrapolating $P_{\mathrm{s}}(L)$ in the thermodynamics limit, i.e., 
$\Delta_{\mathrm{s}}=\lim_{L \rightarrow \infty} \sqrt{P_{\mathrm{s}}(L)/N}$, 
and the result is shown as a function of $U/t$ in Fig.~\ref{fig:Ps_poorman_U}.
The critical point, $U_{\mathrm{c}}/t$, above which the SC order sets
in, and the critical exponent are estimated by assuming a standard
power-law behavior of this quantity close to $U_{\mathrm{c}}$ in the
form of $\Delta_{\mathrm{s}} \sim (U - U_{\mathrm{c}})^{\beta}$, where
$\beta$ is the critical exponent corresponding to the SC order parameter.
Fitting with this form the data of $\Delta_{\mathrm{s}}$ 
evaluated from the simple-minded finite-size scaling,  
we find that there exists a rather large uncertainty, especially in $\beta$. 
Nevertheless, the result indicates that the transition is continuous. 
Therefore, we can apply much more effective and accurate methods to
study the critical behavior.

\begin{figure}[ht]
 \centering
 \includegraphics[width=0.45\textwidth]{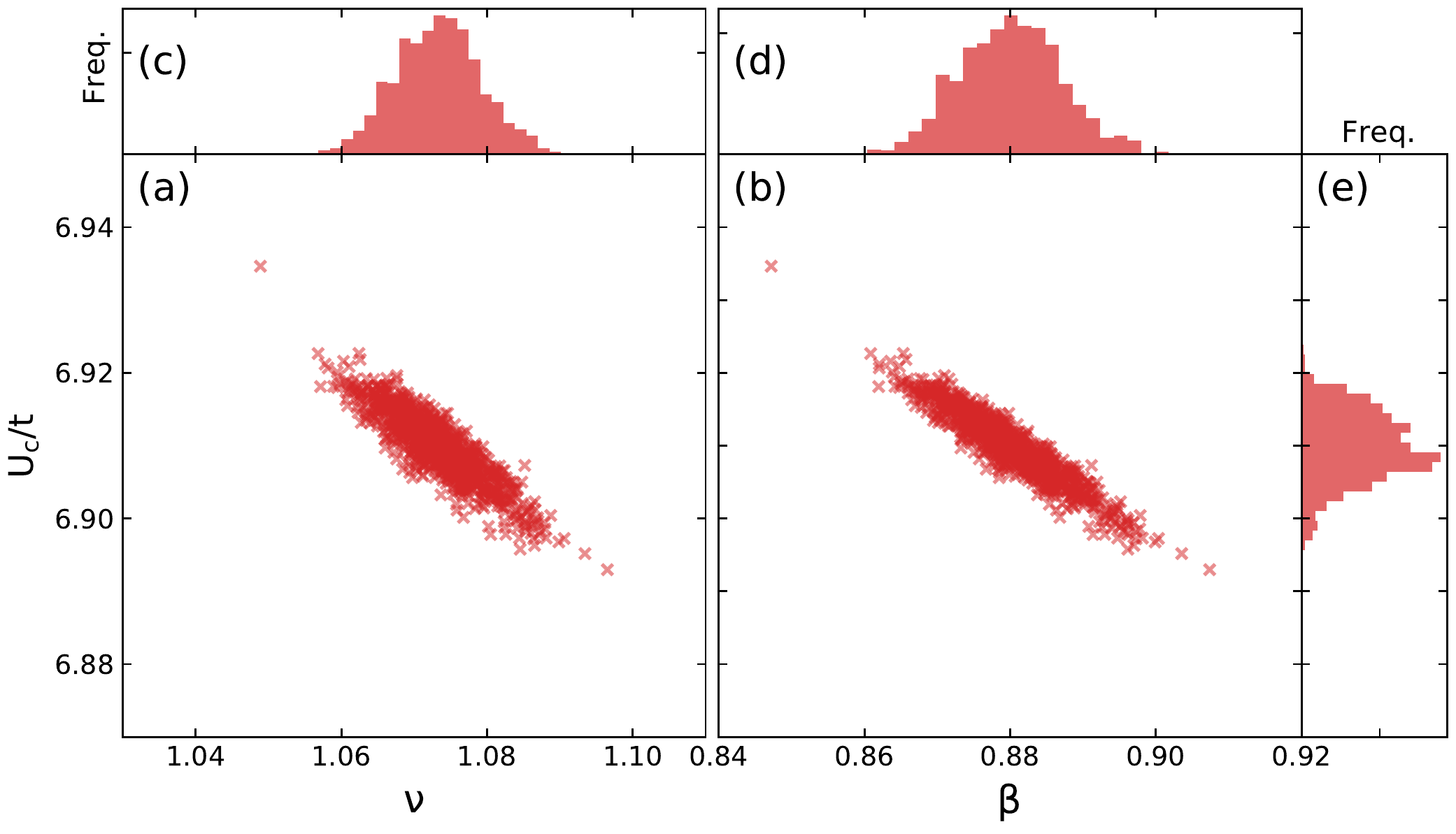}
 \caption{\label{fig:Ps_scatter}%
 (a),(b) Scattering plots and (c)-(e) histograms for the converged values 
 of $U_{\mathrm{c}}/t$, $\nu$, and $\beta$ in the resampling procedure
 for $L_{\mathrm{min}}=18$.
 }
\end{figure}

To this end, we employ the standard method based on the ``collapse fit'' 
that uses all the data points both below and above $U_{\mathrm{c}}/t$. 
This method is known to be very effective for second-order phase transitions. 
For the SC order parameter calculated on each finite-size cluster, 
$\Delta_{\mathrm{s}}(L, U) = \sqrt{P_{\mathrm{s}}(L)/N}$,
we adopt the finite-size scaling relation,
\begin{equation}
 \Delta_{\mathrm{s}}(L, U) = L^{-\beta/\nu} 
f_{\mathrm{s}}(u L^{1/\nu}),
\label{eq:finite-size-scaling_Ps}
\end{equation}
where 
$u=(U-U_{\mathrm{c}})/U_{\mathrm{c}}$ denotes the normalized 
interaction centered at the critical point, and 
$f_{\mathrm{s}}$ represents a scaling function.
According to Eq.~(\ref{eq:finite-size-scaling_Ps}),
the critical points and the exponents are determined by requiring 
that the data points of $\Delta_{\mathrm{s}}(L, U) L^{\beta / \nu}$
as a function of $ u L^{1 / \nu}$ are tightly collapsed to 
a smooth function $f_{\mathrm{s}}$, whose functional form is in general
unknown.
We employ a method based on Bayesian statistics to obtain tight data
collapse in a wide range of $ u L^{1 / \nu}$~\cite{Harada_PRE2011}.
The error bars are evaluated by a resampling technique:
We generate several hundreds of replicas of the QMC data
which are distributed in accordance with the corresponding statistical
errors and repeat the collapse fit for each replica with different
initial parameters of $U_{\mathrm{c}}/t$, $\nu$, and $\beta$.
The error bars are estimated as standard deviations in the distributions 
of the converged parameters~\cite{Otsuka_PRX2016}.
A typical example of this procedure is shown in Fig.~\ref{fig:Ps_scatter}.

\begin{figure}[ht]
 \centering
 \includegraphics[width=0.30\textwidth]{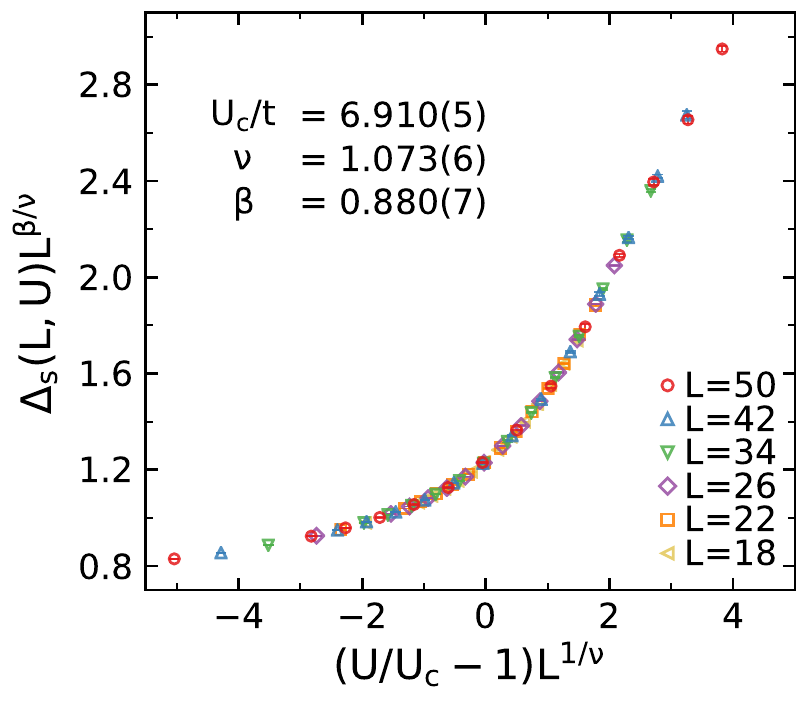}
 \caption{\label{fig:Ps_collapse}%
 Collapse fit of $\Delta_{\mathrm{s}}(L, U)$.
 The critical point and exponents are determined in
 Table~\ref{tbl:exponents_Ps}
 for $L_{\mathrm{min}}=18$.
 }
\end{figure}

The results of the data-collapse analysis are summarized in
Table~\ref{tbl:exponents_Ps}.
We check the finite-size effect by selecting different subsets of 
the data; $L_{\mathrm{min}}$ denotes the smallest
system size used in the data collapse (for example, the results for 
$L=18,22, 26, 34, 50$ are used when $L_{\mathrm{min}}=18$).
In general, one expects that the results converge to the exact values
in the thermodynamics limit with increasing $L_{\mathrm{min}}$.
This is the case notably when the corrections to
the scaling relation are not negligible.
However, since we do not observe a significant drift especially 
in the estimated values of $U_{\mathrm{c}}/t$ with increasing
$L_{\mathrm{min}}$, we expect that our results obtained by the
large-scale simulations have reached the asymptotic scaling regime
governed by  Eq.(\ref{eq:finite-size-scaling_Ps}) without requiring therefore 
the corrections to scaling. 
On the other hand, the statistical errors become large for larger
$L_{\mathrm{min}}$, because the number of data points in the fit
decreases. Thus, as a compromise between reliability and accuracy, 
we set $L_{\mathrm{min}}=18$ to 
estimate the critical indices, and we obtain with this choice a 
well-collapsed fit, as shown in Fig.~\ref{fig:Ps_collapse}.
The values of $U_{\mathrm{c}}/t$ ($\beta$) obtained with this analysis
are smaller (larger) than those estimated from the simple extrapolation
adopted in Fig.~\ref{fig:Ps_poorman_U}.
This is presumably because near the critical point
it is difficult to determine a small value of the order parameter 
by the simple $1/L$ extrapolation~\cite{Assaad_PRX2013}.
We also check that the data points of L$_{\mathrm{4n}}$ in Fig.~\ref{fig:Ps_poorman_L}
are fairly collapsed with these critical exponents.

\begin{table}[ht]
 \caption{\label{tbl:exponents_Ps}%
 Results of the critical point, $U_{\mathrm{c}}/t$, and 
 the exponents, $\nu$ and $\beta$, 
 obtained from data collapse of $\Delta_{\mathrm{s}}(L, U)$.
 $L_{\mathrm{min}}$ denotes the smallest system size 
 used in data collapse. 
 The number in each parenthesis indicates the statistical error, corresponding to the 
 last digit of the value.
 }
 \begin{ruledtabular}
 \begin{tabular}{l l l l }
  $L_{\text{min}}$ & $U_{\mathrm{c}}/t$ & $\nu$ & $\beta$ \\
  \hline
 10 & 6.915(3)   & 1.063(3)   & 0.870(3)   \\
 14 & 6.910(4)   & 1.070(4)   & 0.878(5)   \\
 18 & 6.910(5)   & 1.073(6)   & 0.880(7)   \\
 22 & 6.905(7)   & 1.08(1)    & 0.90(1)  \\
 \end{tabular}
 \end{ruledtabular}
\end{table}

\subsection{\label{subsec:Drmax}quasiparticle weight}

We further examine the values of $U_{\mathrm{c}}/t$ and $\nu$ from
another physical observable in the semimetallic region.
Recently, we have noticed that the quasiparticle weight can be evaluated
from the equal-time single-particle Green's function, which is the basic
quantity and can be calculated with the highest accuracy within the 
the auxiliary-field QMC method~\cite{Seki_InPreparation}.
We consider the equal-time single-particle Green's function,
\begin{equation}
 D(\bm{r}) = \sum_{\sigma} \langle c_{i \sigma}^{\dagger} c_{j \sigma} \rangle,
  \label{eq:Dr}
\end{equation}
where sites $i$ and $j$ belong to different
sublattices, indicated by solid and open circles in Fig.~\ref{fig:lattice}(a), 
and $\bm{r}={\bm r}_i-{\bm r}_j$.
Given that the Green's function near the Fermi level takes 
a Fermi-liquid-type form~\cite{AGD},
it can be shown that 
the Green's function in the long-distance limit is expressed as
\begin{equation}
 \lim_{|\bm{r}| \to \infty}D(\bm{r}) = Z \lim_{|\bm{r}| \to \infty}D^{(0)}(\bm{r}), 
  \label{eq:Z}
\end{equation}
where $Z$ denotes the quasiparticle 
weight~\cite{Migdal_JETP_1957,Luttinger1961,Nozieres1962,Luttinger1962} at the Fermi level 
and $D^{(0)}(\bm{r})$ is the Green's function in the noninteracting limit. 
This relation follows from the fact that the long-distance propagation of a hole is 
determined by the zero-energy excitations at the Dirac points, 
where the amplitude of the hole is renormalized from $1$ in the noninteracting limit to $Z$. 
One may expect that the Fermi velocity is renormalized by the interaction
and thereby have some impact on the Green's function $D(\bm{r})$.
However, it can also be shown that
the renormalization of the Fermi velocity does not change Eq.~(\ref{eq:Z}),
if the linear Dirac dispersion exists, which is expected at least within the semimetallic 
region~\footnote{The noncriticality of the Fermi velocity has been confirmed
in Refs~\onlinecite{Herbut_PRB2009b,Wu_PRB2014,Otsuka_PRX2016}.}.
More detailed discussions on the basis of the repulsive Hubbard model on the honeycomb lattice 
will be published elsewhere~\cite{Seki_InPreparation}.

\begin{figure}[ht]
 \centering
 \includegraphics[width=0.30\textwidth]{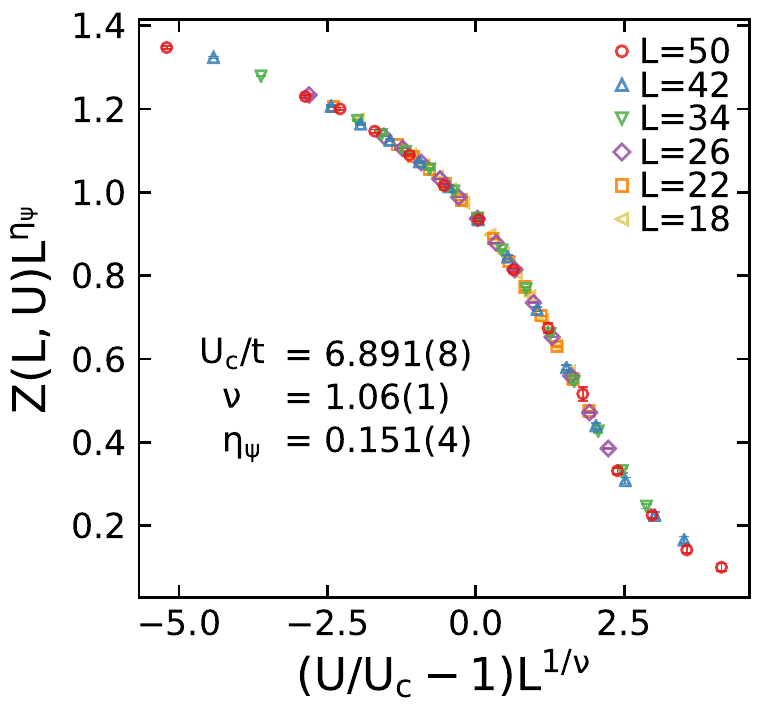}
 \caption{\label{fig:Z_collapse}%
 Collapse fit of $Z(L, U)$. The critical point and exponents 
 are determined in Table~\ref{tbl:exponents_Z} for $L_{\mathrm{min}}=18$. 
 }
\end{figure}

For a finite-size cluster, 
the quasiparticle weight is estimated as
\begin{equation}
 Z(L, U) = \frac{D(\bm{r}_{\mathrm{max}})}{D^{(0)}(\bm{r}_{\mathrm{max}})},
\end{equation}
where the Green's functions are calculated at the maximum distance in
the cluster, $\bm{r}=\bm{r}_{\mathrm{max}}$, so as to detect the
power law behavior at long distance expected in the semimetallic phase.
We perform data-collapse fits for this quantity based on the finite-size 
scaling relation,
\begin{equation}
 Z(L, U) = L^{-\eta_{\psi}} f_{Z}(u L^{1/\nu}),
\label{eq:finite-size-scaling_Z}
\end{equation}
where $\eta_{\psi}$ is the anomalous dimension of the fermion field,
and $f_{Z}$ is a scaling function.
Following the same procedure as in the case of
$\Delta_{\mathrm{s}}(L,U)$, we obtain $U_{\mathrm{c}}/t$, $\nu$, and
$\eta_{\psi}$ in Table~\ref{tbl:exponents_Z}.
It is worth noting that $U_{\mathrm{c}}/t$ and $\nu$ estimated by the 
two independent analyses (Tables~\ref{tbl:exponents_Ps} and~\ref{tbl:exponents_Z})
almost coincide  within two standard deviations.
This is considered to be a nontrivial test for the accuracy of the calculations.
The quality of the data collapse is also excellent, as shown in
Fig.~\ref{fig:Z_collapse}.

\begin{table}[ht]
 \caption{\label{tbl:exponents_Z}%
 Results of the critical point, $U_{\mathrm{c}}/t$, and the exponents,
 $\nu$ and $\eta_{\psi}$, obtained from data collapse of $Z(L, U)$.
 $L_{\mathrm{min}}$ denotes the smallest system size used in data collapse.
 The number in each parenthesis indicates the statistical error, corresponding to the 
 last digit of the value.
 }
 \begin{ruledtabular}
 \begin{tabular}{l l l l l}
  $L_{\text{min}}$ & $U_{\mathrm{c}}/t$ & $\nu$ & $\eta_{\psi}$ \\
  \hline
  10 & 6.913(4)   & 1.036(4)   & 0.162(2)   \\
  14 & 6.896(5)   & 1.056(6)   & 0.154(2)   \\
  18 & 6.891(8)   & 1.06(1)    & 0.151(4)   \\
  22 & 6.89(1)    & 1.05(2)    & 0.154(6)   \\
 \end{tabular}
 \end{ruledtabular}
\end{table}

\section{\label{sec:conclusion}Discussion and Conclusions}

In Table~\ref{tbl:exponents},
we compare our results with those obtained by
the recent QMC simulations~\cite{Li_NatCom2017,Xu_arXiv2018} 
and RG approaches, such as 
the large-$N$ expansion~\cite{Li_NatCom2017,Gracey_PRD2018},
the perturbative RG around the upper critical dimension 
$d=4-\epsilon$~\cite{Rosenstein_PhysLettB1993,Zerf_PRD2017}, and
the functional RG~\cite{Classen_PRB2017}.
Remarkably, our results, both for $\nu$ and $\beta$, agree well
with the QMC results 
on different lattice models which
exhibit the Kekul\'{e} VBS transition~\cite{Li_NatCom2017,Xu_arXiv2018}.
Although the Kekul\'{e} VBS itself has the discrete $Z_{3}$
symmetry breaking instead of $U(1)$, the RG arguments predict an 
emergent $U(1)$ symmetry at the quantum-critical point because of
gapless fermion fluctuations, suggesting that the quantum criticality of
the Kekul\'{e} transition can be described in terms of the chiral XY 
universality class~\cite{Li_NatCom2017,Scherer_PRB2016,Classen_PRB2017}.
The agreement of the critical exponents between the Kekul\'{e} and 
the SC transitions, the latter showing a genuine $U(1)$
symmetry breaking, strongly supports this scenario.
The agreement may also indicate that 
corrections to scaling, which are predicted by the functional RG
argument~\cite{Classen_PRB2017}, turn out to be small
in the Kekul\'{e} VBS transition.

On the other hand, we observe sizable differences between the QMC 
and the analytical estimates, as also previously noticed  in the cases of 
the chiral Heisenberg ~\cite{Otsuka_PRX2016,Zerf_PRD2017}
and the chiral Ising universality classes~\cite{Hesselmann_PRB2016,Zerf_PRD2017}.
However, when compared to the recent higher-order calculations~\cite{Gracey_PRD2018,Zerf_PRD2017}
and the nonperturbative functional RG results~\cite{Classen_PRB2017},
the discrepancies are rather small and systematic.
Our estimate of $\nu$ is always smaller than the analytical results
by $\lesssim$ 10\%.
For $\beta$, the deviations are typically of the order of 20\%.
The most noticeable difference is found in $\eta_{\psi}$. 
For this quantity  only the four-loop RG calculation~\cite{Zerf_PRD2017} yields 
values comparable to our results but still $\lesssim$ 30\% off.
It is clear that
further efforts are required both from the numerical and the analytical approaches
to resolve the remaining discrepancies.

\begin{table}[]
 \caption{\label{tbl:exponents}%
 Critical exponents, $\nu$, $\beta$, and $\eta_{\psi}$, 
 for the chiral XY class.
 QMC results (upper four rows) are based on lattice models,
 while analytical results (shown in the remaining rows)
 are obtained for the GN model.
 The first (second) row represents the results from the analysis of 
 $\Delta_{\mathrm{s}}(L,U)$ [$Z_{\mathrm{s}}(L,U)$] with
 $L_{\mathrm{min}}=18$.
 Two results for the fourth order of the $4-\epsilon$ expansion
 are calculated by different Pad\'{e} approximations.
 The values of $\beta$ which are not directly available in the
 references~\cite{Li_NatCom2017,Xu_arXiv2018,Rosenstein_PhysLettB1993,Zerf_PRD2017,Classen_PRB2017}
 are calculated from the values of the boson anomalous dimension $\eta_{\phi}$
 and the correlation-length exponent $\nu$
 using the scaling relation $\beta=\frac{1}{2}\nu(1+\eta_{\phi})$.
 }
 \begin{ruledtabular}
 \begin{tabular}{l l l l l}
  Method             & 
  $\nu$              & 
  $\beta$            & 
  $\eta_{\psi}$      \\
  \hline
  QMC (present)      & 
  1.073(6)           &
  0.880(7)           &
  $\cdots$           \\
  QMC (present)      &
  1.06(1)            &
  $\cdots$           &
  0.151(4)           \\
  QMC \cite{Li_NatCom2017} &
  1.06(5)            &
  0.90(6)            &
  $\cdots$           \\
  QMC \cite{Xu_arXiv2018} &
  1.05(5)            &
  0.92(5)            &
  $\cdots$           \\
  Large-$N$, 1st order \cite{Li_NatCom2017} &
  1.25               &
  0.75               &
  0.083              \\
  Large-$N$, higher orders \cite{Gracey_PRD2018} &
  1.11               &
  1.05               &
  0.0872             \\
  $4-\epsilon$, 1st order \cite{Rosenstein_PhysLettB1993} &
  0.726              &
  0.619              &
  0.071              \\
  $4-\epsilon$, 2nd order \cite{Rosenstein_PhysLettB1993} &
  0.837              &
  0.705              &
  0.063              \\
  $4-\epsilon$, 4th order \cite{Zerf_PRD2017} & 
  1.19               &
  1.08               &
  0.117              \\
  $4-\epsilon$, 4th order \cite{Zerf_PRD2017} & 
  1.19               &
  1.06               &
  0.108              \\
  functional RG \cite{Classen_PRB2017} & 
  1.16               &
  1.09               &
  0.062              
 \end{tabular}
 \end{ruledtabular}
\end{table}

In conclusion, we have systematically studied the Dirac 
fermions with attractive interaction that allows us to examine the 
$U(1)$ symmetry breaking, i.e., a genuine semimetal-superconductor transition
without the CDW order, occurring at the finite value
of the attractive interaction $U_{\mathrm{c}}$.
By performing large-scale QMC simulations on 
lattices containing up to 2,500 sites, 
we have pinned down the transition from the two different observables, i.e., 
the quasiparticle weight characterizing the semimetal for $U \le U_{\mathrm{c}}$
and the $s$-wave pairing correlation function characterizing the
superconductor for $U \ge U_{\mathrm{c}}$. 
From these quantities, we have obtained 
 consistent values of $U_{\mathrm{c}}$ and  correlation-length
exponent $\nu$. 
Together with the other exponents, $\beta$ and $\eta_{\psi}$, determined also in
this study, our results provide a complete description 
of  the quantum criticality for the chiral XY class in the
GN theory.
Our results represent a useful benchmark calculation for future study of
superconductor phase transitions in condensed matter physics and for an
exhaustive classification of critical phenomena in quantum field theory
models, such as the GN model.

\acknowledgements
We acknowledge
J.~Goryo, Y.~Hatsugai, and S.~Ryu
for valuable comments.
This work has been supported in part by 
Grant-in-Aid for Scientific Research from MEXT Japan (under Grant Nos. 26400413 and 18K03475),
RIKEN iTHES Project, 
MIUR-PRIN-2010. 
The numerical simulations have been performed on K computer 
provided by the RIKEN Center for Computational Science (R-CCS) through 
the HPCI System Research project (Project IDs: hp160159, hp170162, and hp170328),
and RIKEN supercomputer system (HOKUSAI GreatWave).
K. S. acknowledges support from 
JSPS Overseas Research Fellowships. 

\bibliography{chiralXY}
\end{document}